\documentclass[runningheads]{llncs}
\usepackage[utf8]{inputenc} 
\usepackage[T1]{fontenc}    
\usepackage{hyperref}       
\usepackage{cleveref}
\usepackage{url}            
\usepackage{booktabs}       
\usepackage{amsfonts}       
\usepackage{nicefrac}       
\usepackage{microtype}      
\usepackage{graphicx}
\usepackage{todonotes}
\usepackage{csquotes}
\usepackage{siunitx}

\begin{document}
\title{Deep Probabilistic Modeling of \newline Glioma Growth}
\titlerunning{Deep Probabilistic Modeling of Glioma Growth}

\author{
Jens Petersen\inst{1,2,3} \and
Paul F. Jäger\inst{1} \and
Fabian Isensee\inst{1} \and
Simon A. A. Kohl\inst{1\thanks{now with DeepMind and the Karlsruhe Institute of Technology}} \and
Ulf Neuberger\inst{2} \and
Wolfgang Wick\inst{4} \and
Jürgen Debus\inst{5,6,7} \and
Sabine Heiland\inst{2} \and
Martin Bendszus\inst{2} \and
Philipp Kickingereder\inst{2} \and
Klaus H. Maier-Hein\inst{1}
}


\authorrunning{J. Petersen et al.}

\institute{
Div. of Medical Image Computing, German Cancer Research Center, Heidelberg, Germany \and
Dept. of Neuroradiology, Heidelberg University Hospital, Heidelberg, Germany \and
Dept. of Physics \& Astronomy, Heidelberg University, Heidelberg, Germany \and
Dept. of Neurooncology, Heidelberg University Hospital\and
Div. of Molecular and Translational Radiation Oncology, Heidelberg Institute of Radiation Oncology (HIRO), Heidelberg, Germany \and
Heidelberg Ion-Beam Therapy Center (HIT), Heidelberg University Hospital \and
Clinical Cooperation Unit Radiation Oncology, German Cancer Research Center
}




\maketitle

\begin{abstract}
Existing approaches to modeling the dynamics of brain tumor growth, specifically glioma, employ biologically inspired models of cell diffusion, using image data to estimate the associated parameters. In this work, we propose an alternative approach based on recent advances in probabilistic segmentation and representation learning that implicitly learns growth dynamics directly from data without an underlying explicit model. We present evidence that our approach is able to learn a distribution of plausible future tumor appearances conditioned on past observations of the same tumor.
\end{abstract}

\section{Introduction}
\label{sec:introduction}

Glial tumors, especially high grade ones known as glioblastoma multiforme (GBM), are associated with highly irregular growth patterns, involving multiple tissue types for which the change in composition is notoriously difficult to predict. While the prognosis for patients is generally poor, with a median survival of 15-16 months given standard of care treatment for GBM, a considerable number of patients with high grade glioma survive multiple years after diagnosis \cite{bi_beating_2014}. An important factor for this is radiotherapy, which could benefit greatly from a better understanding of growth dynamics.

Most existing approaches that model the growth of glioma do this using variants of the reaction-diffusion equation on the basis of DTI data, e.g. \cite{engwer_glioma_2015,mosayebi_tumor_2012}, but also multi-modal data \cite{menze_generative_2011,le_personalized_2017}, a very recent example being \cite{lipkova_personalized_2019} (for works from before 2011 we refer the reader to \cite{menze_image-based_nodate}). 

In this work, we propose to learn growth dynamics directly from annotated MR image data, without specifying an explicit model, leveraging recent developments in deep generative models \cite{kohl_probabilistic_2018}. We further assume that imaging is ambiguous with respect to the underlying disease (an assumption shared e.g. in \cite{menze_generative_2011,lipkova_personalized_2019}), which is reflected in our approach in that it doesn't predict a single growth estimate (as is done e.g in \cite{akbari_imaging_2016,zhang_convolutional_2018} on a per-pixel level) but instead estimates a distribution of plausible changes for a given tumor. In plain words, we're not interested in the question \enquote{How much will the tumor grow (or shrink)?} but instead \enquote{If the tumor were to grow (or shrink), what would it look like?}. From a clinical perspective, this is relevant for example in radiation therapy, where a margin of possible infiltration around the tumor will also be irradiated. This is currently done in a rather crude fashion by isotropically expanding the tumor's outline \cite{mann_advances_2018}, thus more informed estimates of growth patterns could help spare healthy tissue. Our contributions are the following:

\begin{itemize}
    \item We frame tumor growth modeling as a model-free learning problem, so that all dynamics are inferred directly from data.
    \item We present evidence that our approach learns a distribution of plausible growth trajectories, conditioned on previous observations of the same tumor.
    \item We provide source code: \url{https://github.com/jenspetersen/probabilistic-unet}.
\end{itemize}

\section{Methods}
\label{sec:methods}

The underlying hypothesis of our approach is that tumor growth is at least in part stochastic so that it's not possible to predict a single correct growth trajectory in time from image data alone. Hence, our aim is to \emph{model a distribution of possible changes} of a tumor given the current and in our case one previous observation. We achieve this by training a model to reproduce true samples of observed growth trajectories---with shape and extent of the tumor being represented as multi-class segmentation maps---and using variational inference to allow the model to automatically recognize and account for ambiguity in the task.

\subsection{Data}
\label{sub:data}

We work with an in-house dataset containing a total of 199 longitudinal MRI scans from 38 patients suffering from glioma (15 lower grade glioma and 23 glioblastoma), with a median of 96 days between scans and 5 scans per patient. Patients have undergone different forms of treatment, a fact that we deliberately neglect by declaring it an additional source of ambiguity in the dataset. Each scan consists of 4 contrasts: native T1 (T1n), postcontrast T1 (T1ce), T2 (T2) and fluid-attenuated inversion recovery (FLAIR). All contrasts and time steps for a given patient are skull-stripped, registered to T1 space and resampled to isotropic \SI{1}{mm} resolution \cite{jenkinson_fsl_2012}. For intensity normalization, we only employ basic z-score normalization. Ground truth segmentations of edema, enhancing tumor and necrosis were created semi-automatically by an expert radiologist.

\subsection{Model}
\label{sub:model}

\begin{figure}[h]
\includegraphics[width=\textwidth]{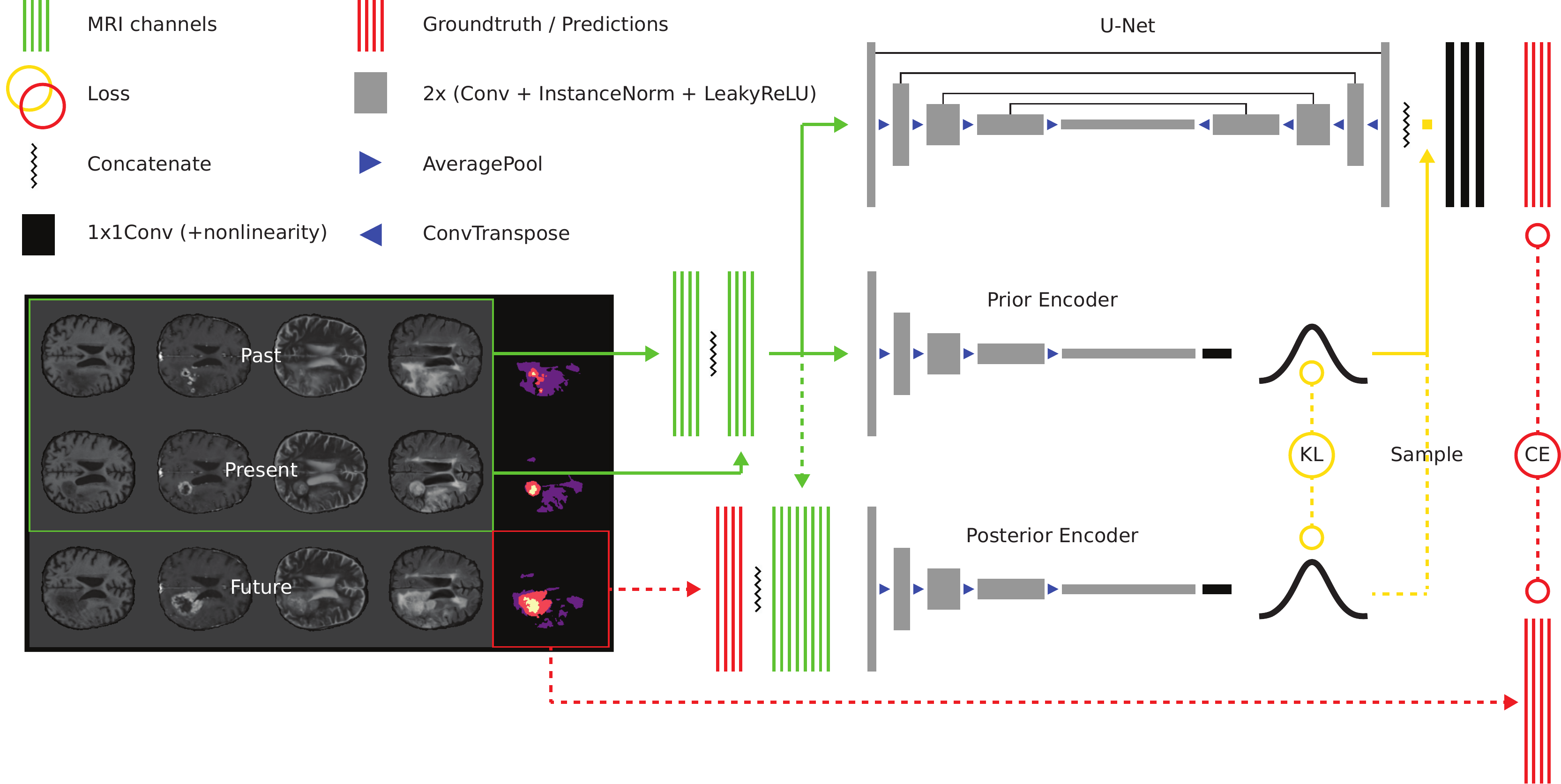}
\caption{The architecture employed in this work. Following the approach in \cite{kohl_probabilistic_2018}, a U-Net \cite{ronneberger_u-net:_2015} is augmented with two additional encoders, one for the prior and one for the posterior. The prior encoder maps the inputs of present and past scans to an $N$-dimensional diagonal Gaussian while the posterior does the same with additional access to the ground truth segmentation from the future. During training, a sample from the posterior is injected into the U-Net after the last decoder block to produce an output that is conditioned on the posterior. During testing the posterior is unavailable and samples can only be drawn from the prior. A KL divergence loss is used to force prior and posterior towards each other while cross entropy is used as segmentation loss. Dashed lines indicate paths that only apply during training.}
\label{fig:architecture}
\end{figure}

Our model along with the training procedure, based on a probabilistic segmentation approach \cite{kohl_probabilistic_2018}, are visualized in \cref{fig:architecture}. The architecture comprises three components: 1. A U-Net \cite{ronneberger_u-net:_2015} to map scans from present and past to future tumor appearance. 2. A fully convolutional encoder that maps scans from present and past to an $N$-dimensional diagonal Gaussian (the \emph{prior}; we choose $N=3$). 3. An encoder with the same architecture that maps scans from present and past as well as the ground truth segmentation from the future to another diagonal Gaussian (the \emph{posterior}; $N=3$). During training we sample from the posterior and concatenate the sample to the activations of the last decoder block in the U-Net, so as to condition the softmax predictions on the sample. We employ multi-class cross entropy as the segmentation loss and use the Kullback-Leiber divergence to force prior and posterior towards each other, so that at test time---when a ground truth segmentation is no longer available---the predicted prior is as close as possible to the unknown posterior. This objective is the well known \emph{evidence lower bound} used in variational inference.


The described training scheme will give rise to the following desirable properties: 1) The model will learn to represent the task's intrinsic ambiguity in the Gaussian latent space, in our case different plausible future tumor shapes and sizes, as we show in \cref{sec:results}. 2) At test time we can sample multiple consistent hypotheses from the latent space (as seen in \cref{fig:latent}), and select those that match desired criteria (e.g. tumor volume increases by $20\%$).

We train with data augmentation (using \url{https://github.com/MIC-DKFZ/batchgenerators}) on patches of size $112^3$, but evaluate on full sized scans of $192^3$. For details on optimization and associated hyperparameters we refer to the provided source code (\url{https://github.com/jenspetersen/probabilistic-unet}).

\subsection{Experiments \&\ Evaluation}
\label{sub:experiment}

We seek to show that our approach learns meaningful future tumor appearances instead of just segmentation variants of the present input. For this reason we construct a baseline that is restricted to learning the latter.

Let A denote past, B present and C future. Our model is trained and evaluated for triples $AB\to C$ (that we will refer to as cases), as shown in \cref{fig:architecture}. An upper bound on performance is given by a regular probabilistic segmentation model that is trained and evaluated with tuples $C\to C$. This is essentially a model that has complete knowledge of the future and just needs to segment it. At the same time, a model trained on $B\to B$ but evaluated on $B\to C$ can serve as a lower bound to our model trained on $AB\to C$. If our performance matches that of the lower bound, we have learned to produce plausible segmentations for the current time step, but not the future.

We split our subjects randomly into 5 groups and perform 5-fold cross validation, i.e. we train on 4 subsets and predict the remaining one. For many triples, the real change between time steps is small, which makes it hard to show that our approach actually learns meaningful change. As a consequence we define two groups to report results for:

\begin{enumerate}
    \item\textbf{Large Change}: The 10\% of cases with the most pronounced change in terms of whole tumor Dice overlap, resulting in a threshold of 0.48 and 13 cases.
    \item\textbf{Moderate Change} The cases with larger than mean change (0.70), but not in top 10\%, resulting in 44 cases.
\end{enumerate}

We are not interested in predictive capabilities, so it makes little sense to look at the overlap of the prior mean predictions with the future ground truth (our approach performs not much better than the lower bound here). We report metrics that are representative of our model's desired capabilities, 1) a clinically relevant question, i.e. what the tumor will look like for a given expected size, and 2) how well the model is able to represent large changes in its latent space: 

\begin{enumerate}
    \item \textbf{Query Volume Dice}: We take samples from a grid around the prior mean ($-3\sigma$ to $+3\sigma$ in steps of $1\sigma$) and select the segmentation for which the whole tumor volume (i.e. all tumor classes contribute) best matches that of the ground truth. If our approach is able to model future appearances, it should perform better than the lower bound with increasing real change.
    \item \textbf{Surprise}: This is the KL divergence the model assigns for a given combination of past \&\ present scans and future ground truth. A lower KL divergence between prior and posterior means the model deems the combination more realistic, i.e. it is less \emph{surprised}.
\end{enumerate}

\section{Results}
\label{sec:results}

\subsection{Qualitative Results}
\label{sub:qualitative}

We first present several qualitative examples, selected to illustrate the types of changes our approach is able to represent.

\begin{figure}
    \centering
    \includegraphics[width=0.8\textwidth]{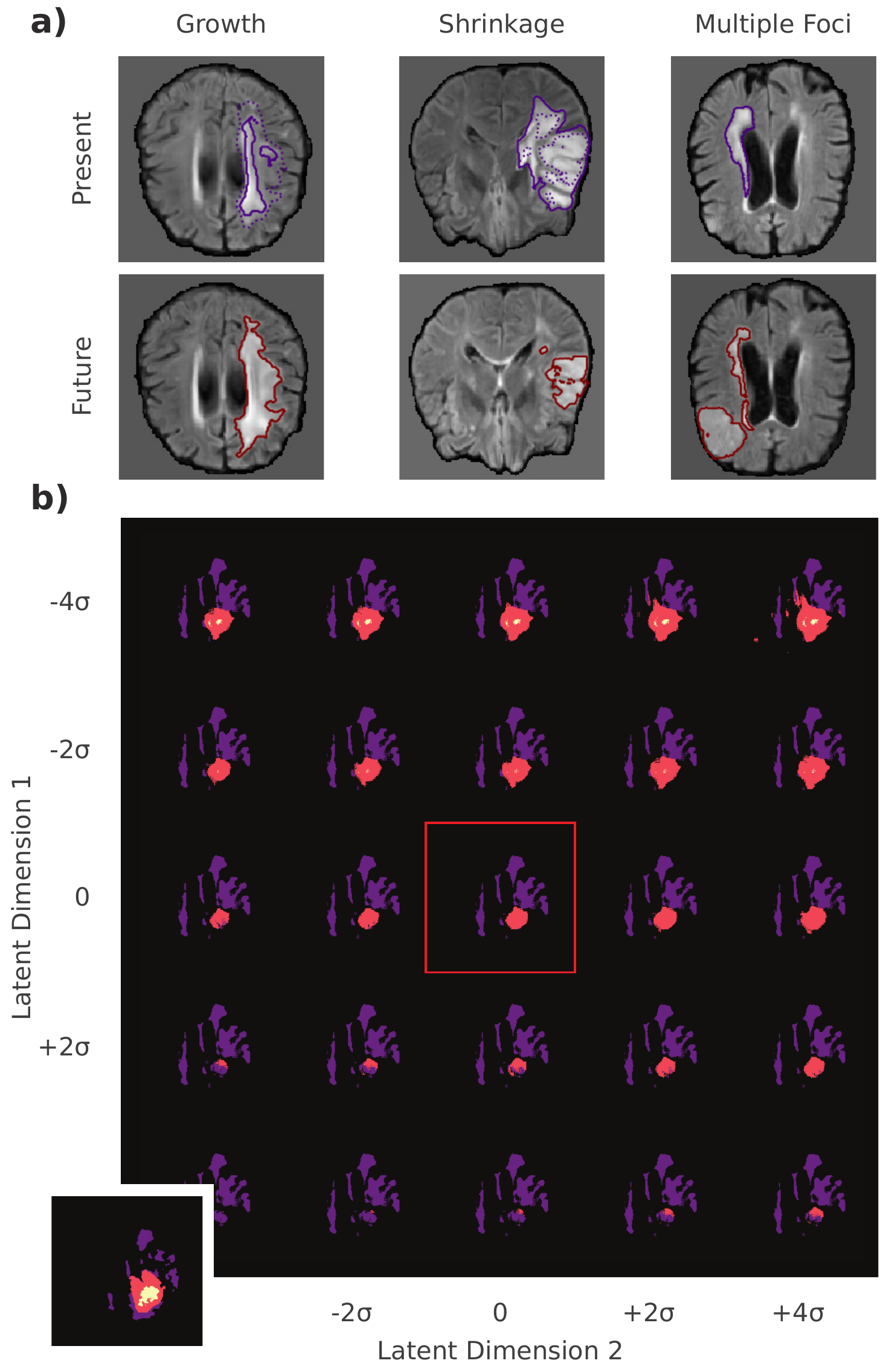}
    \caption{Qualitative Examples:
  (a) Prior mean prediction (solid purple) and sample with best volume match (dashed purple) as well as future ground truth (red) overlaid on FLAIR. The approach is able to model growth or shrinkage, but is unable to represent tumors with both growth and shrinkage in different locations (for multiple foci, dotted and solid overlap). (b) Regular grid samples from prior, with mean highlighted in red and ground truth inlay in bottom left corner (unrelated to (a)). The learned latent space separates class contributions, dimension 1 seems to encode tumor core size (enhancing tumor and necrosis) while dimension 2 encodes enhancing tumor size (note how necrosis is virtually constant in the top row). The third latent dimension, not shown here, captures small variations in edema size. Purple -- Edema, Orange -- Enhancing Tumor, Yellow -- Necrosis}
    \label{fig:latent}
\end{figure}

\Cref{fig:latent} a) shows three cases with outlines for the prior mean (solid purple) prediction as well as the sample from the prior (dotted purple) that best matches the volume of the real future (red). The similarity of the latter two in the first two columns indicates that our model is able to represent both strong growth and strong reduction in size well. It can also be seen that the mean prediction closely matches the current state of the tumor, which is unsurprising, because small changes occur most frequently. The third column is illustrative of a general limitation of our model: encoding into the latent space removes all spatial resolution, so tumors that both shrink and grow in different locations (e.g. with multiple foci) are not represented in the current setup.

 \Cref{fig:latent} b) illustrates how the learned latent space represents semantically meaningful continuous variations: Dimension 2 changes the size of just the enhancing tumor while dimension 1 changes the size of the tumor core (enhancing tumor and necrosis combined). The third axis that is not shown encodes variation in the size of the edema, meaning that the model automatically learned to separate the contributions of the different tumor regions. Most importantly, all variations seem plausible. Note that while a reduction in necrosis is biologically implausible in a treatment-naive context, it might very well occur under treatment like in our dataset.

\subsection{Quantitative Results}
\label{sub:quantitative}

In this section we compare our approach with an upper bound and a lower bound. These are given by a regular probabilistic U-Net \cite{kohl_probabilistic_2018} trained for segmentation with (upper bound) and without (lower bound) knowledge of the future and both evaluated with respect to future ground truth.

\begin{figure}[h]
    \centering
    \includegraphics[width=\textwidth]{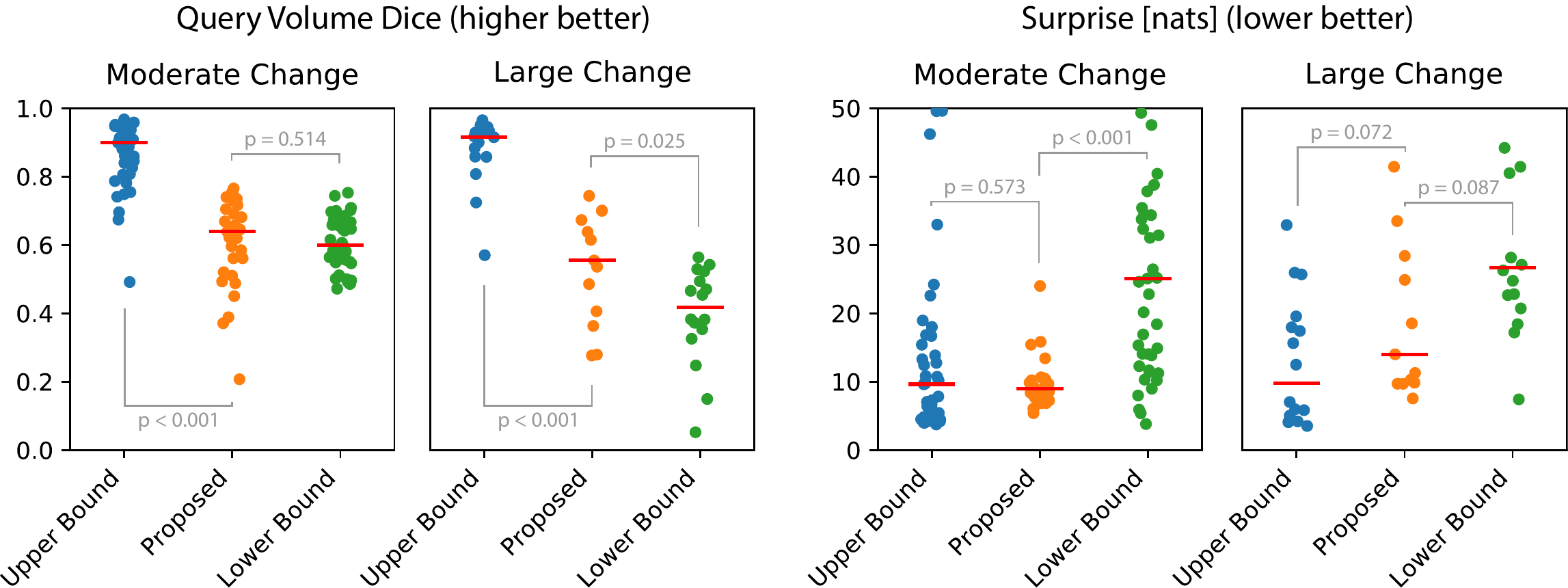}
    \caption{Quantitative results for \emph{Query Volume Dice} and \emph{Surprise}, for groups with moderate and large change and median indicated in red, p-values from Wilcoxon rank-sum test. For large changes, our approach can represent the future much better than the lower bound. The low surprise in our model indicates that our model's learned prior assigns higher likelihood than the lower bound to the real future tumor appearance, leveraging temporal information from previous scans.}
    \label{fig:scores}
\end{figure}

\Cref{fig:scores} shows median results for two different metrics and both moderate change and large change. \emph{Query Volume Dice} represents the clinically motivated question of estimating spatial extent for a given change in size (e.g. for radiation therapy). Particularly for cases with large change our approach outperforms the lower bound. At the same time, the \emph{Surprise}, a measure of how close estimated prior and posterior are for a given set of inputs and future ground truth, is on par with the upper bound for cases with moderate change and still much lower than the lower bound's for large change cases. For reference, in VAEs this usually comes at the cost of poor reconstruction, but the reconstruction loss (i.e. segmentation cross entropy, not shown) is also much lower for our approach compared to the lower bound in both cases.

\section{Discussion}

In this work we investigated whether glioma growth dynamics can be learned directly from data without an underlying explicit biological model, instead relying on probabilistic segmentation to model distributions of future tumor appearances.

Our results indicate that this is indeed possible. We showed quantitatively that our approach can represent large variations in the inferred distributions and that these learned distributions model growth trajectories instead of just segmentation variants for a known input. Qualitative examples show overall realistic growth as well as shrinkage patterns. Compared to existing work, our approach relies on a very different hypothesis, so we elected to present metrics that evaluate our desired goals, but are unfortunately unsuitable for quantitative comparison with classical methods.

Our work has a number of shortcomings we'd like to explicitly address. While our dataset is larger than what is usually presented in the literature on glioma growth, our method clearly requires more data than existing ones that are based on explicit biological diffusion models. Without a doubt the dataset is too small to be representative of all plausible growth variations. We were also unable to apply our approach on more than two input time steps, because this reduced the amount of available training instances too drastically. As we pointed out, our model is also unable (and not designed) to predict a single correct growth trajectory. It is further unable to resolve spatially varying growth for a single tumor, likely because we employ a simple global latent space. On the other hand, more complex models would again require more data. Finally, it would be desirable to represent time continuously instead of working with discrete steps.

Contrasting the above, we also see some advantages that our approach offers. The ability to sample consistent hypotheses from the latent space, as opposed to just having pixel-wise probability estimates, lends itself to answering clinically motivated questions, e.g. exploring only samples that correspond to strong growth or those that produce predictions where a certain region is or is not affected by the tumor. We further don't rely on imaging modalities like DTI that are not typically acquired in clinical routine. It would in fact be interesting to explore if our approach can benefit from including the latter.

Overall we feel our work opens up a promising new avenue of approaching glioma growth and tumor growth in general. We are confident that much larger datasets will become available in the future that will allow our method to further improve. Most importantly, our work is entirely complementary with respect to diffusion-based models, and combining them should be exciting to explore. 

\bibliographystyle{splncs04}
\bibliography{1242-refs}

\end{document}